\documentclass[12pt,twoside]{article}
\usepackage{fleqn,espcrc1,epsfig}

\usepackage{graphicx}
\usepackage[figuresright]{rotating}

\newcommand{\AmS}{{\protect\the\textfont2
  A\kern-.1667em\lower.5ex\hbox{M}\kern-.125emS}}

\title{Abnormal radioactive decays out of long-lived\\ super- and
 hyper-deformed isomeric states\footnote{Contribution presented at the NS2000
 conference,
 East Lansing, Michigan, August 15 - 19, 2000.(Version 1 is a shorter write-up
 of a previous talk presented in the ENS 2000 conference, Debrecen,
  Hungary, May 15 - 20, 2000).}}

\author{A. Marinov\address{Racah Institute of
Physics, The Hebrew University, Jerusalem 91904, Israel},
S.Gelberg$^a$
        and
        D. Kolb\address{Department of Physics, University GH Kassel,
34109 Kassel, Germany}}

\begin{document}

\maketitle \vspace*{-0.2cm}
\begin{abstract}
 Long-lived high-spin super- and hyper-deformed isomeric
states, which exhibit themselves by abnormal radioactive decays,
have been observed using the $^{16}$O + $^{197}$Au and $^{28}$Si +
$^{181}$Ta reactions. They make it possible to understand the
production, via secondary reactions, of the long-lived superheavy
element with Z = 112 and of the abnormally low energy and very
enhanced $\alpha$-particle groups seen in various actinide
sources. They might also explain some puzzling phenomena seen in
nature.
\end{abstract}
\vspace*{-0.2cm}
\section{INTRODUCTION}
\vspace*{-0.2cm}
 In the past several observations were made which
could not be understood under the given knowledge of nuclear
physics. First \cite{mar71,mar84}, evidence for the production,
via secondary reactions in CERN W targets, of a long-lived
superheavy isotope $^{\sim272}$112 has been obtained. Fission
fragments were observed in Hg sources chemically separated from
the W targets, and the measured masses undergoing fission (like
308, 315 and 318) were consistently interpreted \cite{mar84} as
due to 5 different common molecules (like -Cl, -N$_{3}$ and
-NO$_{2}$) of a superheavy isotope with Z=112 (Eka-Hg),
N$\sim$160. There was however no understanding of the long deduced
half-life of several weeks and of the large deduced fusion cross
section of several mb \cite{mar84,bar92}. In a study of actinide
fractions from the same W target, long-lived isomeric states were
found in the neutron-deficient $^{236}$Am and $^{236}$Bk nuclei
with half-lives of 0.6~y and $\geq$30 d, respectively
\cite{mar87}. Their character was not clear, being far from closed
shell nuclei, where high spin isomers are known, and living much
longer than the known fission isomers. In addition, several
unidentified $\alpha$-particle groups were found in some actinide
sources. Thus, 5.14 MeV (t$_{1/2}$ = 3.8$\pm$1 y),
  5.27 MeV  (t$_{1/2}$
= 625$\pm$84 d) and  5.53 MeV (t$_{1/2}$ = 26$\pm$7 d) groups were
respectively found in the Bk, Es and Lr-No sources
\cite{mar87,mar78,mar87y,mar99,mar00b} (See fig. 1). Again, one
could not understand their relatively low energies (e.g., 5.53 MeV
in Lr-No as compared to  g.s. to g.s. transitions around  8 MeV,
which have about 13 orders of magnitude larger penetrability
factors), and very short half-lives (10$^{5}$ - 10$^{7}$ shorter
than predicted from
 the systematics \cite{vio66}).
 The deduced evaporation-residue cross sections
\cite{mar00b}, in the mb region, are also several orders larger
than expected.

A number of unexplained phenomena have  also been observed in
nature. Po halos, produced by $\alpha$-particles from $^{210}$Po
(t$_{1/2}$ = 138.4 d), $^{214}$Po (t$_{1/2}$ = 164 $\mu$s) and
$^{218}$Po (t$_{1/2}$ = 3.0 m), have been seen in mica
\cite{hen39,gen68,gen88}. Hulubei and Cauchois \cite{hul47} saw
induced Po X-rays in the mineral petzite (Ag$_{3}$AuTe$_{2}$). In
both cases the precursors Th and U were not present.

\vspace*{-1.0cm}
\begin{figure}[h]
\includegraphics[width=1.0\textwidth,angle={-1.0}]{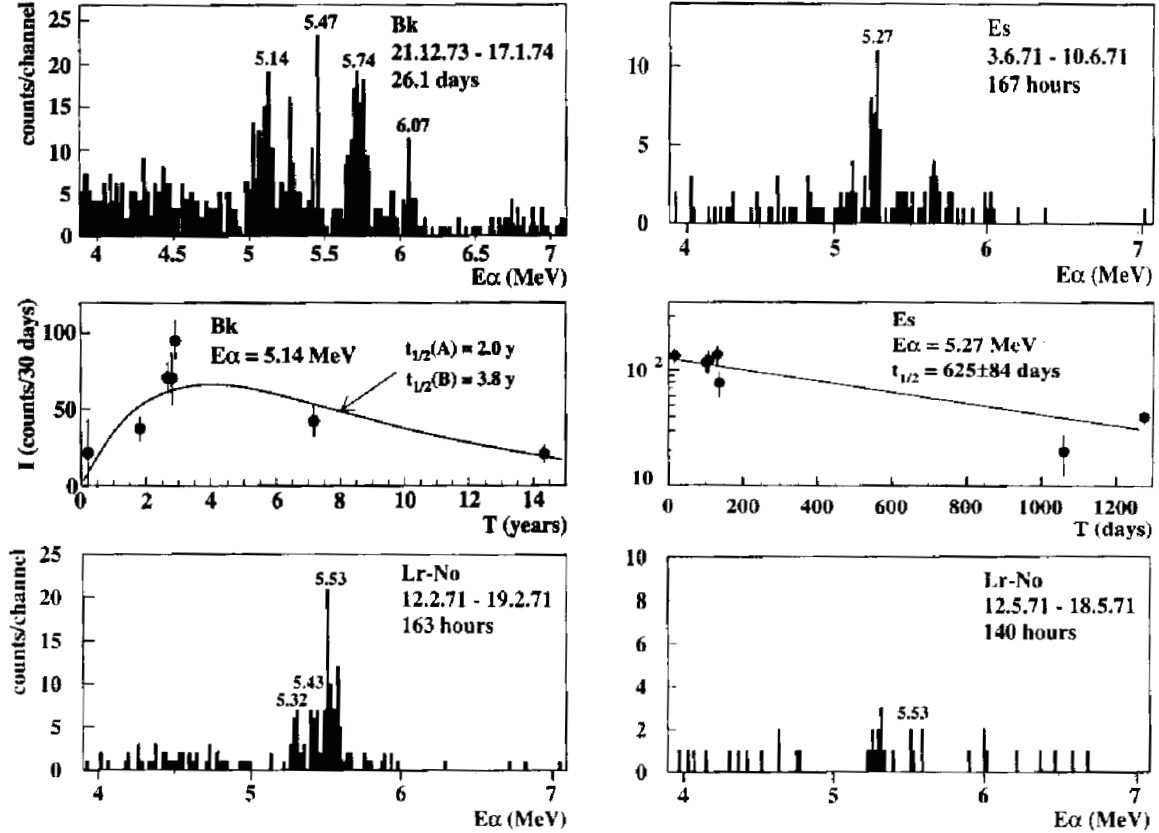}
\vspace*{-1.5cm} \caption{{\bf Left, top:} $\alpha$-particle
spectrum obtained with the Bk source. {\bf Right, top:}
$\alpha$-particle spectrum obtained with the Es source. {\bf Left,
center:} Decay curve obtained with the 5.14 MeV group seen with
the Bk source (left, top above).(See comment (c) in table~1
regarding the growing half-life of 2.0 y [8]). {\bf Right,
center:} Decay curve obtained with the 5.27 MeV group seen with
the Es source (right, top above). {\bf Left, bottom:}
$\alpha$-particle spectrum obtained with the Lr-No source. {\bf
Right, bottom:} The same as the previous one but taken about 3
months later. From a comparison of the two spectra a half-life of
26$\pm$7 d was deduced for the 5.53 MeV group.}
\end{figure}

\vspace*{-0.3cm}
 Evidence for unidentified $\alpha$-particles around  4.5
MeV have been seen in several minerals
\cite{schi38,cher64,che68,mei70}. According to the chemistry used
they were thought \cite{che68} to be due to Eka-Os (element 108).
It was also suggested \cite{mei70} that $^{247}$Cm may be a
descendant of this superheavy element. These results were not
substantially convincing. The predicted \cite{vio66} half-life for
such low energy $\alpha$-particles  in Z = 108 is around 10$^{16}$
y, implying the existence of an impossibly  large amount of
material of about 0.1 g in the studied samples.

In the following a consistent possible interpretation for these
phenomena will be given, based on the recently discovered
\cite{mar96a,mar96b,mar00a} long-lived high-spin super- and
hyper-deformed isomeric states.

%

\section{SUPER- AND HYPER-DEFORMED ISOMERIC STATES\\
 IN THE
  $^{16}$O + $^{197}$Au AND $^{28}$Si +
$^{181}$Ta REACTIONS} \vspace*{-0.2cm}
   The $^{16}$O + $^{197}$Au and  $^{28}$Si + $^{181}$Ta reactions
  have been studied using  beams from the Pelletron accelerator
  in Rehovot, catcher foil technique  and long-period off-line measurements
  in Jerusalem.
        In  the first
reaction at E$_{Lab}$ = 80 MeV \cite{mar96a,mar96b}, long-lived
high-spin super-deformed isomeric states have been found. They
exhibited themselves by decaying with relatively low energy (5.2
MeV) and very enhanced (t$_{1/2}$ $\sim$ 90 m, 3x10$^{5}$
enhancement) $\alpha$-particles  in coincidence with $\gamma$-rays
of a superdeformed band (SDB), and also
by long-lived proton decays with half-lives of about 6 and 70 h.
In  the $^{28}$Si + $^{181}$Ta reaction at E$_{Lab}$ = 125 MeV
\cite{mar99,mar00a}, about 10\% below the Coulomb barrier, a high
energy (7.8 - 8.6 MeV) and long-lived (40~d $\leq$ t$_{1/2}$
$\leq$ 2.1 y) $\alpha$-particle group was found in coincidence
with SDB $\gamma$-rays (fig. 2), and consistently interpreted as
due to production of  a high-spin long-lived hyper-deformed
isomeric state in $^{195}$Hg, which decays by a factor of
10$^{13}$ retarded $\alpha$-particles to SDB states in $^{191}$Pt.
($^{195}$Hg may be produced via 1p1n evaporation reaction and 3
consecutive $\alpha$-decays. The probability that almost all the
measured $\gamma$-ray energies will accidentally fit the predicted
energies of a SDB transitions is 4.1x10$^{-8}$. (The background is
zero). The high energy $\gamma$-rays are most probably sum events
rather than single photo-peak events. Therefore the number of
Compton events is relatively low [20]).

\vspace*{-0.8cm}
\begin{figure}[h]
\includegraphics[width=0.45\textwidth]{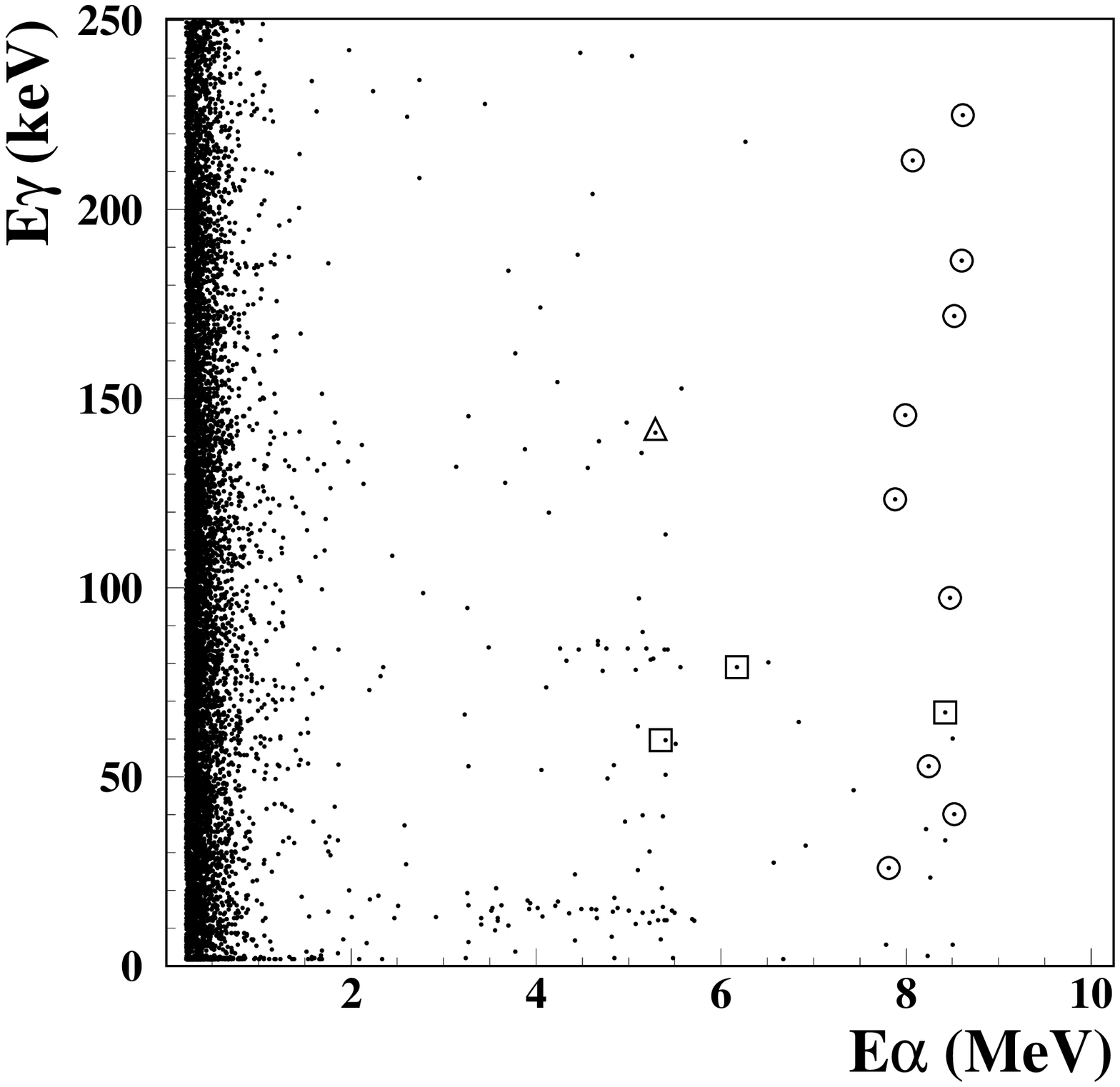}
\includegraphics[width=0.41\textwidth]{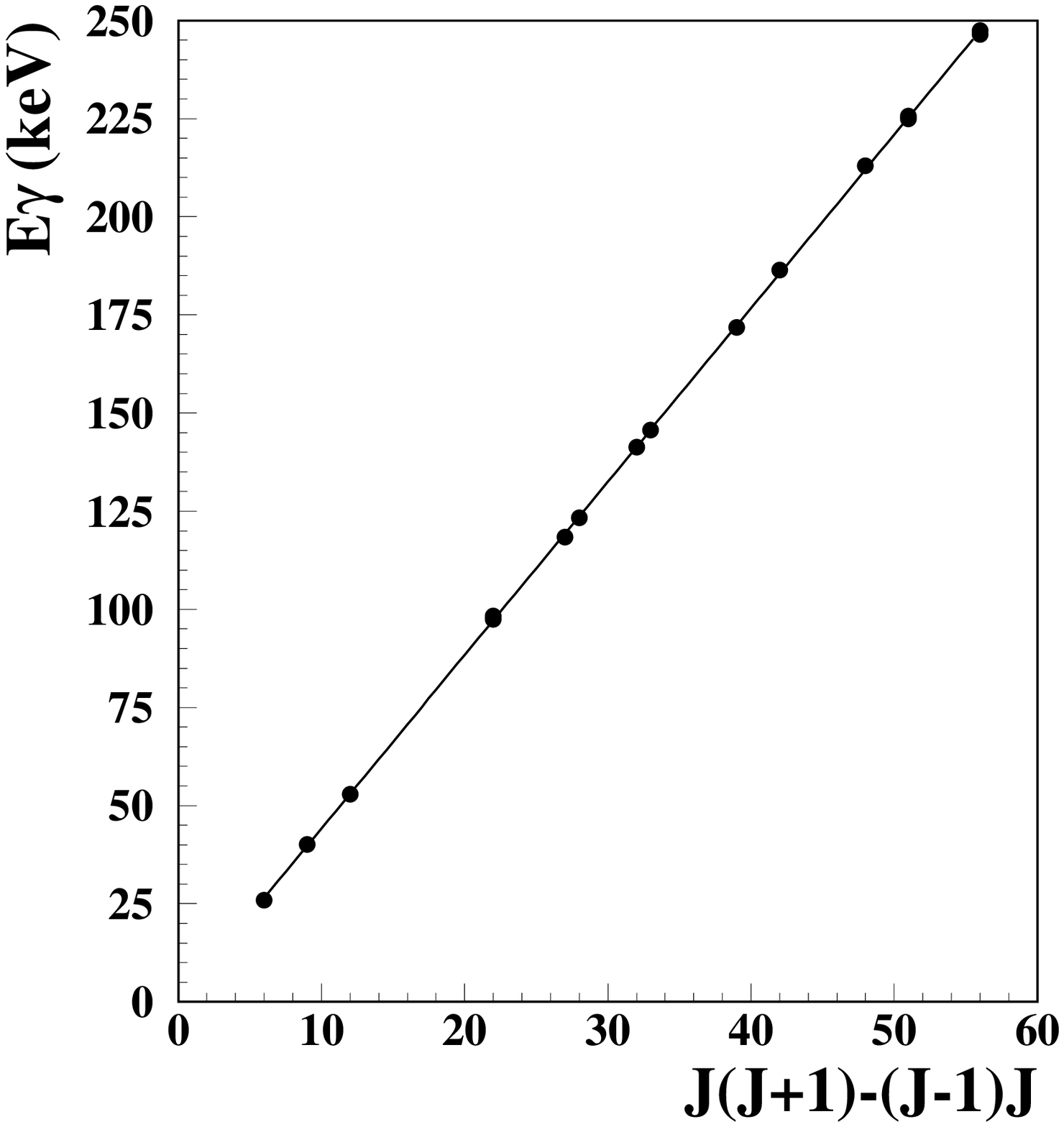}
\vspace*{-0.9cm}
 \caption{{\bf Left:}
$\alpha$ - $\gamma$ coincidence plot from one measurement of the
$^{28}$Si + $^{181}$Ta reaction. E$_{Lab}$ = 125 MeV, with 200
$\mu$g/cm$^{2}$ C catcher foil, taken for 76.8 d, starting 77.4 d
after the end of irradiation. The $\gamma$ - ray energies of the
encircled events fit with SDB transitions. The squared events fit
with known characteristic X - rays and the events in triangles are
identified with known $\gamma$ - ray transitions (see ref. [20]).
{\bf Right:} E$\gamma$ versus J(J+1)-(J-1)J for the $\gamma$ -
rays seen in coincidence with 7.8 - 8.6 MeV $\alpha$ - particles.
 (The encircled events in the left figure plus similar
events obtained in a second measurement [20]).}
\end{figure}

\vspace*{-1.1cm}
\section{SUPER- AND HYPER-DEFORMED ISOMERIC STATES\\ IN THE ACTINIDES}
\vspace*{-0.2cm}
 Based on these results and on the predicted
\cite{how80} excitation energies of the second and  third  minima
in various actinide nuclei, the unidentified $\alpha$-particle
groups from the actinide fractions (Section 1) can be consistently
interpreted \cite{mar99,mar00b},
 both from the low
energy and the enhanced lifetime points of view, as due to
II$^{min}$ $\rightarrow$ II$^{min}$ or III$^{min}$ $\rightarrow$
III$^{min}$ transitions (see table 1). The 5.14 MeV group in Bk
grew at the beginning (fig.~1, left, center). It therefore may be
due to a II$^{min}$ $\rightarrow$ II$^{min}$ transition from
$^{238}$Am or a III$^{min}$ $\rightarrow$ III$^{min}$ transition
from $^{238}$Cm, which were produced from $^{238}$Bk by the
EC/$\beta$$^{+}$ process \cite{mar00b,mar87}. (Fig. 6 of ref.
\cite{mar00b} indicates that the first case, where a rather deep
second minimum is predicted \cite{how80}, is more likely). The
5.27 and 5.53 MeV groups in the Es and Lr-No sources are
consistent with III$^{min}$ $\rightarrow$ III$^{min}$ transitions
from $^{247}$Es and $^{252}$No, respectively.

\begin{table}[h]
\vspace*{-0.5cm} \caption[]{Predicted $\alpha$-particle energies
[21]  and half-lives for various transitions between
super-deformed and hyper-deformed minima in Am - No nuclei. The
values in {\bf bold} are consistent with the experimental results
mentioned in Section 1. }
\begin{minipage}{1\textwidth} 
\renewcommand{\footnoterule}{\kern -3pt} 
\begin{tabular}{llllll}
\hline\\[-10pt]         
Mother &     E$\alpha$(MeV) &     E$\alpha$(MeV)$^{a}$ &
E$\alpha$(MeV)$^{a}$
 &
t$_{1/2}$$^{b}$ ($\beta$$_{2}$; $\beta$$_{3}$; $\beta$$_{4}$)  &
t$_{1/2}^{exp}$\\
  Isotope & g.s.$\rightarrow$g.s. & II$^{min}$$\rightarrow$II$^{min}$
   & III$^{min}$$\rightarrow$III$^{min}$\\
\hline\\[-10pt]
 $^{238}$Am$^{c}$  &    5.94   &   {\bf5.13} &  4.53 &
10.8 y (0.71; 0.0; 0.09) & 3.8$\pm$1 y\\
 $^{238}$Cm$^{c}$   &  6.51  &
5.90  &  {\bf 5.24} &   0.5 y (1.05; 0.17; 0.0) & 3.8$\pm$1 y\\
 $^{247}$Es &  7.37 & 7.47 & {\bf5.27} & 382 d (1.05; 0.19; 0.0) & 625$\pm$84 d\\
 $^{252}$No  &  8.42 &  $\sim$7.8 & $\sim${\bf5.6} &
  81 d (1.20; 0.19; 0.0) & 26$\pm$7 d\\
\hline
\end{tabular}
\end{minipage}
\renewcommand{\footnoterule}{\kern-3pt \hrule width .4\columnwidth
\kern 2.6pt}            
\end{table}
\vspace*{-1.0cm}

$^{a}$ Deduced from ref. [21].

$^{b}$ Calculated according to formulas given in refs. [8,18].
$\beta$$_{2}$ and $\beta$$_{4}$ values were deduced from the
$\epsilon$$_{2}$ and $\epsilon$$_{4}$ values given in ref. [21]
using fig. 2 from W. Nazarewicz and I. Ragnarsson, Handbook of
Nuclear Properties. eds. D. Poenaru and W. Greiner (Clarendon
Press, Oxford, 1996) p. 80. The value of $\beta$$_{3}$ was taken
equal to $\epsilon$$_{3}$.

$^{c}$ Since the intensity of the 5.14 MeV group  in  Bk  grew at
the beginning (fig. 1, left, center), it may be due to a decay of
an isotope of Cm or Am which is produced from Bk by the
EC/$\beta^{+}$ process. (Similarly to the $^{236}$Bk and
$^{236}$Am cases which decay to $^{236}$Pu [4]).\\

\begin{minipage}[b] {0.46\linewidth}


\centering\epsfig{figure=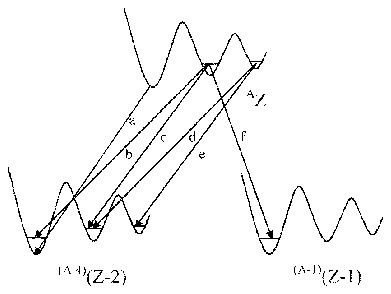,width=1.1\linewidth}
\vspace*{0.5cm}

\end{minipage}\hfill
\begin{minipage}[b] {0.46\linewidth}
 a) I$^{min}$$\rightarrow$I$^{min}$. Normal
$\alpha$'s.\\
 b) II$^{min}$$\rightarrow$I$^{min}$. Retarded $\alpha$'s:\\
 \hspace*{2.0cm} $^{190}$Ir$\rightarrow$$^{186}$Re (Ref. [20]).\\ c)
II$^{min}$$\rightarrow$II$^{min}$. Enhanced $\alpha$'s:\\
 \hspace*{2.0cm} $^{210}$Fr$\rightarrow$$^{206}$At (Ref. [18]).\\
 \hspace*{1.0cm} $\sim$$^{238}$Am$\rightarrow$$^{234}$Np (Ref. [8]).\\ d)
III$^{min}$$\rightarrow$II$^{min}$. Retarded $\alpha$'s:\\
 \hspace*{2.0cm} $^{195}$Hg$\rightarrow$$^{191}$Pt (Ref. [20]).\\ e)
III$^{min}$$\rightarrow$III$^{min}$. Enhanced $\alpha$'s:\\
 \hspace*{2.0cm} $^{\sim247}$Es$\rightarrow$$^{243}$Bk (Ref.
 [8]).\\
 \hspace*{2.0cm} $\sim$$^{252}$No$\rightarrow$$^{248}$Fm (Ref. [8]).\\ f)
II$^{min}$$\rightarrow$I$^{min}$. Retarded protons:\\
 \hspace*{2.0cm} $^{198}$Tl$\rightarrow$$^{197}$Hg(?) (Ref.
 [19]).\\
 \hspace*{2.0cm} $^{205}$Fr$\rightarrow$$^{204}$Rn(?) (Ref. [20]).\\
\end{minipage}
 Figure 3. Summary of abnormal particle decays
seen in various experiments.\\

The long-lived isomeric states in the $^{236}$Am and $^{236}$Bk
nuclei can be interpreted as due to the existence of super- or
hyper-deformed  isomers in these nuclei which decay by the EC or
$\beta^+$ processes eventually to the normal states in $^{236}$Pu
\cite{mar87}.

Fig. 3 summarizes the various new and abnormal particle decays
which were observed in different experiments. It should be
mentioned that a high-spin long-lived super-deformed isomeric
state has been predicted back in 1969 \cite{nil69}.

 \vspace*{-0.3cm}
\section{THE LONG-LIVED SUPERHEAVY ELEMENT WITH Z = 112}
\vspace{-0.2cm}
 The experiments \cite{mar71,mar84} concerning
element 112 can now be understood as due to its production in a
long-lived super- or hyper-deformed isomeric state, rather than in
the ground state. The large deduced fusion cross section is due to
two effects. a) Very little extra-push energy is needed in order
to produce the compound nucleus in such a state. b) The fusion
probability in the secondary reactions, where the projectile is
itself a deformed fragment produced just within 2x10$^{-14}$ s
before interacting with another W nucleus in the target, is much
larger as compared to fusion  with a normal projectile, due to the
much reduced Coulomb repulsion \cite{mar98}.
 \vspace*{-0.3cm}
\section{ISOMERIC STATES AND SUPERHEAVY ELEMENTS IN NATURE?}
\vspace{-0.2cm}
 The discovered high spin super- and hyper-deformed
isomeric states may be the source for the Po halos
\cite{hen39,gen68,gen88}  (either in the $^{210}$Po, $^{214}$Po
and $^{218}$Po isotopes which decay by $\gamma$'s to the  ground
states, or in the appropriate Pb or Bi nuclides  which decay by
$\beta$$^{-}$ to the corresponding Po isotopes), and also for the
observed Po induced X-rays in petzite \cite{hul47}.
\vspace*{-0.5cm}
\begin{figure}[h]
\begin{center}
\includegraphics[width=0.55\textwidth]{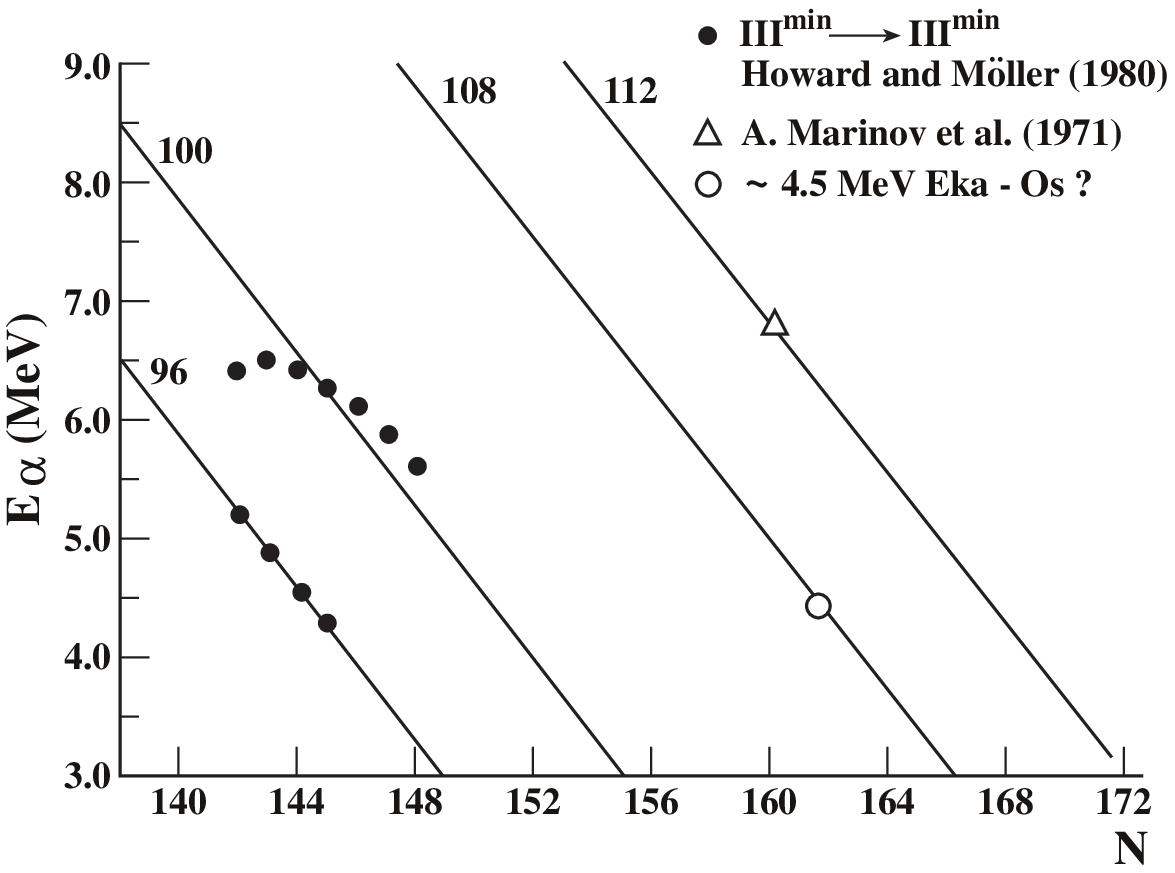}
\end{center}
\end{figure}

\vspace*{-1.1cm} Figure 4. Predictions \cite{how80}, and
extrapolations from these predictions, of the III$^{min}$
$\rightarrow$ III$^{min}$ $\alpha$-particle energies. The black
dots are the predictions for various isotopes of Z=96 and Z=100.
The straight lines are extrapolations from these predictions for
Z=108 and Z=112. The open triangle shows the position of the 6.73
MeV $\alpha$-particles seen with the Hg sources from the CERN W
targets \cite{mar71}. The open circle shows the position of 4.5
MeV $\alpha$-particles in Z=108.

\vspace*{0.3cm} The low energy $\alpha$-particles around 4.5 MeV
\cite{schi38,cher64,che68,mei70} may consistently be interpreted
as due to a very enhanced III$^{min}$ $\rightarrow$ III$^{min}$
transition in Z$\sim$108 and A$\sim$271. The predicted
\cite{mar96a,mar00b}  half-life in this case is around 10$^{9}$ y,
which implies an amount of material of about 10 ng. This resolves
the main difficulty in understanding these data where an
impossible amount of 0.1 g was deduced assuming a normal
$\alpha$-transition (see the Introduction). In addition, an
extrapolation of the predictions \cite{how80} of $\alpha$-energies
of
 III$^{min}$ $\rightarrow$ III$^{min}$ transitions
(see fig. 4), shows that for Z=108, E$\alpha$ of about 4.5 MeV
corresponds to N$\sim$162. This is consistent with the suggestion
\cite{mei70} that $^{247}$Cm may be a descendent of the superheavy
element with Z=108 which decays by the 4.5 MeV $\alpha$-particles,
since $^{247}$Cm can be obtained from $^{271}_{108}$Hs$_{163}$ by
successive 6 $\alpha$-decays. It should however be mentioned that
in principle the above  4.5 MeV $\alpha$-particles may  also be
due to  a very retarded II$^{min}$ $\rightarrow$ I$^{min}$ or
III$^{min}$ $\rightarrow$ II$^{min}$ transition in the region of
Os itself. (For normal 4.5 MeV $\alpha$-particles in Os the
expected \cite{vio66} half-life is about 1 y. Such short-lived
nuclide can not exist in nature).

\vspace{0.2cm}
 We would like to thank J. L. Weil for taking part
in the measurements with the actinide sources, A. Pape for
bringing the work of ref. \cite{hul47} to our attention and N.
Zeldes for very valuable discussions.\\

\end{document}